\documentclass[twocolumn,english,reprint, longbibliography, superscriptaddress, breaklinks=true, showkeys, showpacs=false, nofootinbib]{revtex4}
\usepackage[T1]{fontenc}
\usepackage[latin9]{inputenc}
\usepackage{color}
\usepackage{babel}
\usepackage{amsmath}
\usepackage{amsthm}
\usepackage{amsfonts}
\usepackage{amssymb}
\usepackage{graphicx}
\usepackage{subfigure}
\usepackage{physics}

\usepackage[unicode=true,pdfusetitle,
 bookmarks=true,bookmarksnumbered=false,bookmarksopen=false,
 breaklinks=true,pdfborder={0 0 0},backref=false,colorlinks=true]
 {hyperref} 
\usepackage{breakurl}

\makeatletter
\@ifundefined{textcolor}{}
{%
 \definecolor{BLACK}{gray}{0}
 \definecolor{WHITE}{gray}{1}
 \definecolor{RED}{rgb}{1,0,0}
 \definecolor{GREEN}{rgb}{0,1,0}
 \definecolor{BLUE}{rgb}{0,0,1}
 \definecolor{CYAN}{cmyk}{1,0,0,0}
 \definecolor{MAGENTA}{cmyk}{0,1,0,0}
 \definecolor{YELLOW}{cmyk}{0,0,1,0}
}

\pdfoutput=1
\hypersetup{colorlinks=true,citecolor=blue,linkcolor=cyan,urlcolor=blue,filecolor= green, breaklinks=true}
\usepackage{url}
\usepackage{breakurl}
\makeatother

\begin{document}

\title{Interferometric Visibility in Curved Spacetimes}

\author{Marcos L. W. Basso}
\email{marcoslwbasso@mail.ufsm.br}
\address{Departamento de F\'isica, Centro de Ci\^encias Naturais e Exatas, Universidade Federal de Santa Maria, Avenida Roraima 1000, Santa Maria, Rio Grande do Sul, 97105-900, Brazil}

\author{Jonas Maziero}
\email{jonas.maziero@ufsm.br}
\address{Departamento de F\'isica, Centro de Ci\^encias Naturais e Exatas, Universidade Federal de Santa Maria, Avenida Roraima 1000, Santa Maria, Rio Grande do Sul, 97105-900, Brazil}

\selectlanguage{english}%

\begin{abstract} 
 In [M. Zych \textit{et al.}, Nat. Commun. 2, 505 (2011)],  the authors predicted that the interferometric visibility is affected by a gravitational field in way that cannot be explained without the general relativistic notion of proper time. In this work, we take a different route and start deriving the same effect using the unitary representation of the local Lorentz transformation in the Newtonian Limit. In addition, we show that the effect on the interferometric visibility due to gravity persists in different spacetime geometries.  However, the influence is not necessarily due to the notion of proper time. For instance, by constructing a `astronomical' Mach-Zehnder interferometer in the Schwarzschild spacetime, the influence on the interferometric visibility can be due to another general relativistic effect, the geodetic precession. Besides, by using the unitary representation of the local Lorentz transformation,  we show that this behavior of the interferometric visibility is general for an arbitrary spacetime, provided that we restrict the motion of the quanton to a two-dimensional spacial plane.
\end{abstract}

\keywords{Interferometric Visibility; Curved spacetimes; General relativistic effects}

\maketitle

\section{Introduction}

Recently, there has been an increasing effort for probing the interplay between gravity and quantum mechanics or, more specifically, to witness the quantumness of gravity \cite{Bose, Chiara, Carlo, Howl}, as well as to probe general relativistic effects in quantum phenomena \cite{Futamase, Zych, Magdalena, Brodutch, Costa}. For instance, in Ref. \cite{Zych} the authors predicted a quantum effect that cannot be explained without the general relativistic notion of proper time. They considered a Mach-Zehnder interferometer placed in a gravitational potential, with a `clock' used as an interfering particle, i.e., an evolving internal degree of freedom of a particle. Due to the difference in proper time elapsed along the two trajectories, the `clock' evolves to different quantum states for each path of the interferometer. Because of the wave-particle duality, the interferometric visibility will decrease by an amount given by the which-way information accessible from the final state of the clock, which gets entangled with the external degree of freedom of the particle.

Moreover, concern about how entanglement behaves in relativistic scenarios has grown more and more \cite{Terno}. For example, the authors of Refs. \cite{Adami, Ueda} showed that the entanglement of Bell states depends on the velocity of the observer. On the other hand, the authors in Ref. \cite{Milburn} argued that the overall entanglement of a Bell state remains invariant for a Lorentz boosted observer, the entanglement is just shuffled between the different degrees of freedom. However, it was demonstrated by Peres \textit{et al.} \cite{Peres} that the entropy of a single massive spin-1/2 particle does not remain invariant under Lorentz boosts. These apparently conflicting results involve systems containing different particle states and boost geometries \cite{Palge}. Therefore, entanglement under Lorentz boosts is highly dependent on the boost scenario in question \cite{Dunningham}. More generally, the entanglement for observers uniformly accelerated in a flat space-time was considered in Refs. \cite{Alsing, Schuller, Fuentes}. A step forward in the investigations of these relativistic scenarios was taken by Terashima and Ueda \cite{Terashima}, who studied EPR correlations and the violation of Bell's inequalities in curved spacetimes, by considering a succession of infinitesimal local Lorentz transformations. In addition, the same authors, using the same method in Ref. \cite{Ueda1}, studied the decoherence of spin states due to the presence of a gravitational field.

In this work, based on the method developed in Ref. \cite{Terashima}, we start deriving the same effect of Ref. \cite{Zych}. We use the unitary representation of the local Lorentz transformation in the Newtonian Limit. In addition, as we will show here, the effect on the interferometric visibility due to gravity persists in different spacetime geometries. However, the oscillation is not necessarily due to the difference of the proper time elapsed in each path. For instance, by constructing a `astronomical' Mach-Zehnder interferometer in the Schwarzschild spacetime, the influence on the interferometric visibility is due to another general relativistic effect, i.e., geodetic precession. Besides, we show that by using the unitary representation of the local Lorentz transformation, this behavior of the interferometric visibility in an arbitrary spacetime is general, provided that we restrict the motion of the quanton to a two-dimensional spacial plane. The benefit of the approach taken here, through the representation of the local Lorentz transformation, is that one does not need to know the (internal) Hamiltonian of the system and how it couples to the gravitation field, as in Refs. \cite{Zych, Costa}. In contrast, given an arbitrary spacetime metric, we only need to calculate the Wigner rotation, which is a straightforward procedure.

The organization of this article is as follows. In Sec. \ref{sec:spin}, we review the spin-$1/2$ dynamics in curved spacetimes. In Sec. \ref{sec:inter}, we study the behavior of interferometric visibility in different spacetime geometries. Thereafter, in Sec. \ref{sec:con}, we give our conclusions.

%-------------------------------
\section{Spin Dynamics in Curved Spacetimes}
\label{sec:spin}
\subsection{Spin States in Local Frames }
The study of the dynamics of spin-$1/2$ particles in gravitational fields requires the use of local frames of reference defined at each point of spacetime. These frames are defined through an orthonormal basis or tetrad field (or vielbein), which is a set of four linearly independent orthonormal 4-vector fields \cite{Wald}. The differential structure of the spacetime, which is a differential manifold $\mathcal{M}$ \cite{Carroll}, provides, in each point $p$, a coordinate basis for the tangent space $T_{p}(\mathcal{M})$, as well as for the cotangent space $T^*_{p}(\mathcal{M})$, given by $\{\partial_{\mu}\}$ and $\{ dx^{\nu} \}$, respectively, such that $dx^{\nu}(\partial_{\mu}) := \partial_{\mu}x^{\nu} = \delta^{\ \nu}_{\mu}$. Therefore, the metric can be expressed as $g = g_{\mu \nu}(x) dx^{\mu} \otimes dx^{\nu}$, and the elements of the metric, which encodes the gravitational field, are given by $g_{\mu \nu}(x) = g(\partial_\mu, \partial_{\nu})$. Since the coordinate basis $\{\partial_{\mu}\} \subset T_p(\mathcal{M})$ and $\{ dx^{\nu} \} \subset T^*_p(\mathcal{M})$ are not necessarily orthonormal, it is always possible set up any basis as we like. In particular, we can form an orthonormal basis with respect to the pseudo-Riemannian manifold (spacetime) on which we are working. Following Ref. \cite{Nakahara}, let us consider the linear combination
\begin{align}
    & e_a = e_a^{\ \mu}(x) \partial_\mu, \ \ \ e^a = e^a_{\ \mu}(x)dx^{\mu}, \\
    & \partial_{\mu} = e^a_{\ \mu}(x)e_a, \ \ \ dx^{\mu} = e_a^{\ \mu}(x)e^a.
\end{align}  
To define a local frame at each point $p \in \mathcal{M}$, we require $\{e_a\}$ to be orthonormal in the following sense
\begin{align}
    g(e_a, e_b) := \eta_{ab}, \ \ \ g := \eta_{ab}e^a \otimes e^b,
\end{align}
where $\eta_{ab} = diag(-1,1,1,1)$ is the Minkowski metric. Equivalently, we can define the tetrad field in terms of its components
\begin{align}
    & g_{\mu \nu}(x)e_a^{\ \mu}(x)e_b^{\ \nu}(x) = \eta_{ab},\\ &  \eta_{ab}e^a_{\ \mu}(x)e^b_{\ \nu}(x) = g_{\mu \nu}(x) \label{eq:metr},
\end{align}
with
\begin{align}
    e^a_{\ \mu}(x)e_b^{\ \mu}(x) = \delta^{a}_{\ b}, \ \ \ e^a_{\ \mu}(x)e_a^{\ \nu}(x) = \delta_{\mu}^{\ \nu}.
\end{align}
In what follows, Latin letters $a, b, c, d,\cdots$ refer
to coordinates in the local frame; Greek indices $\mu, \nu, \cdots$ run over the four general-coordinate labels; and repeated indices are to be summed over. The components of the tetrad field and its inverse transforms a tensor in the general coordinate system into one in the local frame, and vice versa. Therefore it can be used to shift the dependence of spacetime curvature of the tensor fields to the tetrad fields. In addition, Eq. (\ref{eq:metr}) informs us that the tetrad field encodes all the information about the spacetime curvature hidden in the metric. Besides, the tetrad field $\{e_a^{\ \mu}(x), a = 0,1,2,3\}$ is a set of four 4-vector fields, which transforms under local Lorentz transformations in the local system. The choice of the local frame is not unique, since the local frames remains local under the local Lorentz transformations. Therefore, a tetrad representation of a particular metric is not uniquely defined, and different tetrad fields will provide the same metric tensor, as long as they are related by local Lorentz transformations \cite{Misner}.

By constructing the local Lorentz transformation, we can define a particle with spin-$1/2$ in curved spacetimes as a particle whose one-particle states furnish the spin-$1/2$ representation of the local Lorentz transformation \cite{Terashima}. Thus, let's consider a massive spin-$1/2$ particle moving with four-momentum $p^{\mu}(x) = m u^{\mu}(x)$ with $p^{\mu}(x) p_{\mu}(x) = -m^2$, where $m$ is the mass of the quanton, $u^{\mu}(x)$ is the four-velocity in the general coordinate system, and we already put $c = 1$. Now, we can use the tetrad field $e^a_{\ \mu}(x)$ to project the four-momentum $p^{\mu}(x)$ into the local frame, i.e., $p^a(x) = e^a_{\ \mu}(x) p^{\mu}(x)$. Thus, in the local frame at point $p \in \mathcal{M}$ with coordinates $x^a = e^a_{\ \mu}(x) x^{\mu}$, a momentum eigenstate of a Dirac particle in a curved spacetime is given by \cite{Lanzagorta}
\begin{align}
 \ket{p^a(x), \sigma; x} := \ket{p^a(x), \sigma; x^{a}, e^a_{\ \mu}(x), g_{\mu \nu}(x)},
\end{align}
and represents the state with spin $\sigma$ and momentum $p^a(x)$ as observed from the position $x^a = e^a_{\ \mu}(x) x^{\mu}$ of the local frame defined by $ e^a_{\ \mu}(x)$ in the spacetime $\mathcal{M}$ with metric $g_{\mu \nu}(x)$. 

The description of a Dirac particle state can only be provided regarding the tetrad field and the local structure that it describes, since $e^0_{\ \mu}$ produces a preferred global time coordinate from the local time coordinate. By definition, the state $\ket{p^a(x), \sigma; x}$ transforms as the spin-$1/2$ representation under the local Lorentz transformation. In the case of special relativity, a one-particle spin-$1/2$ state $\ket{p^a, \sigma}$ transforms under a Lorentz transformation $\Lambda^{a}_{b}$ as \cite{Weinberg}
\begin{equation}
    U(\Lambda)\ket{p^a, \sigma} = \sum_{\lambda} D_{\lambda \sigma} (W(\Lambda,p)) \ket{\Lambda p^a, \lambda},
\end{equation}
where $D_{\lambda, \sigma}(W(\Lambda,p))$ is a unitary representation of the Wigner's little group, whose elements are Wigner rotations $W^{a}_{b} (\Lambda,p)$ \cite{Eugene}. The subscripts can be suppressed and one can write $U(\Lambda) \ket{p^a, \sigma} = \ket{\Lambda p^a} \otimes D (W(\Lambda, p)) \ket{\sigma},$ as sometimes we'll do. In other words, under a Lorentz transformation $\Lambda$, the momenta $p^a$ goes to $\Lambda p^a$, and the spin transforms under the representation $D_{\sigma, \lambda}(\Lambda, p)$ of the Wigner's little group \cite{Onuki}. Meanwhile, in a curved spacetime everything above remains essentially the same, except by the fact that single-particle states now form a local representation of the inhomogeneous Lorentz group at each point $p \in \mathcal{M}$, i.e., 
\begin{equation}
    U(\Lambda(x))\ket{p^a(x), \sigma;x} = \sum_{\lambda} D_{\lambda \sigma}(W(x)) \ket{\Lambda p^a(x), \lambda;x} \label{eq:unit},
\end{equation}
where $W(x) := W(\Lambda(x), p(x))$ is a local Wigner rotation.

%-------------------------
\subsection{Spin Dynamics}
Following Terashima and Ueda \cite{Terashima}, let us consider how the spin changes when the quanton moves from one point to another in a curved spacetime. In the local  frame at point $p$ with coordinates $x^a = e^a_{\ \mu}(x) x^{\mu}$, the momentum of the particle is given by $p^a(x) = e^a_{\ \mu}(x) p^{\mu}(x)$. After an infinitesimal proper time $d \tau$, the quanton moves to a new point with general coordinates $x'^{\mu} = x^{\mu} + u^{\mu} d\tau$. Then, the momentum of the particle in the local frame at the new point becomes $p^a(x') = p^a(x) + \delta p^a(x)$, where the variation of the momentum in the local frame can be described by the combination of changes due to non-gravitational external interaction $\delta p^{\mu}(x)$, and spacetime geometry effects $\delta e^a_{\ \mu}(x)$:
\begin{equation}
    \delta p^a(x) = e^a_{\ \mu}(x) \delta p^{\mu}(x) + \delta e^a_{\ \mu}(x)p^{\mu}(x).
\end{equation}
The variation $\delta p^{\mu}(x)$ in the first term on the right hand side of the last equation is simply given by
\begin{equation}
     \delta p^{\mu}(x) = u^{\nu}(x) \nabla_{\nu} p^{\mu}(x) d\tau = m a^{\mu}(x) d\tau, \label{eq:momen}
\end{equation}
where $\nabla_{\nu}$ is the covariant derivative and $a^{\mu}(x):=u^{\nu}(x) \nabla_{\nu} u^{\mu}(x)$ is the acceleration due to a non-gravitational interaction. Once $p^{\mu}(x)p_{\mu}(x) = -m^2$ and $p^{\mu}(x)a_{\mu}(x) = 0$, Eq. (\ref{eq:momen}) can be rewritten as
\begin{equation}
     \delta p^{\mu}(x) = - \frac{1}{m}(a^{\mu}(x)p_{\nu}(x) - p^{\mu}(x)a_{\nu}(x))p^{\nu}(x) d\tau.
\end{equation}
Meanwhile, the variation of the tetrad field is given by
\begin{align}
    \delta e^a_{\ \mu}(x) & = u^{\nu}(x) \nabla_{\nu}e^a_{\ \mu}(x) d\tau \nonumber \\
    & = - u^{\nu}(x) \omega_{\nu \ b}^{\ a}(x) e^b_{\ \mu}(x)d \tau,
\end{align}
where $\omega_{\nu \ b}^{\ a} := e^{a}_{\ \lambda} \nabla_{\nu} e_{b}^{\ \lambda} = - e_{b}^{\ \lambda} \nabla_{\nu} e^{a}_{\ \lambda} $ is the connection 1-form (or spin connection) \cite{Chandra}. Collecting these results and substituting in Eq. (\ref{eq:momen}), we obtain
\begin{equation}
    \delta p^a(x) = \lambda^{a}_{\ b}(x)p^{b}(x) d\tau \label{eq:momvar}, 
\end{equation}
where
\begin{align}
    \lambda^{a}_{\ b}(x) & = - \frac{1}{m}(a^{a}(x)p_{b}(x) - p^{a}(x)a_{b}(x)) + \chi^{a}_{\ b} \nonumber \\
    & = - (a^{a}(x)u_{b}(x) - u^{a}(x)a_{b}(x)) + \chi^{a}_{\ b} \label{eq:infloc}
\end{align}
with $\chi^{a}_{\ b} :=  - u^{\nu}(x) \omega_{\nu \ b}^{\ a}(x)$. It can be shown that Eqs. (\ref{eq:momvar}) and (\ref{eq:infloc}) constitute an infinitesimal local Lorentz transformation since, as the particle moves in spacetime, the momentum in the local frame will transform under an infinitesimal local Lorentz transformation $p^{a}(x) = \Lambda^{a}_{\ b}(x) p^b(x)$ where $\Lambda^{a}_{\ b}(x) = \delta^{a}_{\ b} + \lambda^{a}_{\ b}(x)d \tau$ \cite{Lanzagorta}. If the particle moves in a geodesic in spacetime, then $a^{\mu}(x) = 0$ and the infinitesimal Lorentz transformation in the local frame reduces to $\lambda^{a}_{\ b}(x) = -u^{\nu}(x) \omega_{\nu \ b}^{\ a}(x)$. Classicaly, we can define the spin by a four-vector $s^{\mu}$ which is orthogonal to the momentum \cite{Ryder}, $s^{\mu}p_{\mu} = 0$. In this case, the spin will be affected according to $u^{\nu}\nabla_{\nu} s^{\mu} = 0$, i.e., the spin is parallel transported along the geodesic, provided that we neglect the spin-curvature coupling. It is well known that, due to the curvature effects, the spin will not return to its initial state after being parallel transported along an arbitrary closed path, even when the path of the particle corresponds to a geodesic. Moreover, since geodesic paths preserve the magnitude of vectors, the spin magnitude remains constant along the path, hence the change of the spin state is associated with the spin precession. This phenomena is known as geodetic precession \cite{Misner}, and, quantum mechanically, can be associated with the action of successive local Wigner rotations \cite{Salgado}.

Now, given the local Lorentz transformation, we can construct the local Wigner rotation that affects the spin of the particle. In other words, by using a unitary representation of the local Lorentz transformation, the state $\ket{p^a(x), \sigma; x}$ is now described as $U(\Lambda(x)) \ket{p^a(x), \sigma; x}$ in the local frame at the point $x'^{\mu}$, and Eq. (\ref{eq:unit}) expresses how the spin of the quanton rotates locally as the particle moves from $x^{\mu} \to x'^{\mu}$ along its world line. Therefore, one can see that spacetime tells quantum states how to evolve. For the infinitesimal Lorentz transformation, the infinitesimal Wigner rotation is given by
\begin{equation}
    W^{a}_{\ b}(x) = \delta^{a}_{\ b} + \vartheta^{a}_{\ b} d \tau,
\end{equation}
where $\vartheta^{0}_{\ 0}(x) = \vartheta^{i}_{\ 0}(x) = \vartheta^{0}_{\ i}(x) = 0$ and
\begin{equation}
    \vartheta^{i}_{\ j}(x) = \lambda^{i}_{\ j}(x) + \frac{\lambda^{i}_{\ 0}(x)p_j(x) - \lambda_{j0}(x)p^i(x)}{p^0(x) + m}.
\end{equation}
In Ref. \cite{Kilian}, the authors provided an explicit calculation of these elements, and the two-spinor representation of the infinitesimal Wigner rotation is then given by
\begin{align}
    D(W(x)) & = I_{2 \times 2} + \frac{i}{4} \sum_{i,j,k = 1}^{3} \epsilon_{ijk} \vartheta_{ij}(x) \sigma_k d \tau \nonumber \\
    & = I_{2 \times 2} + \frac{i}{2} \boldsymbol{\vartheta} \cdot \boldsymbol{\sigma} d\tau, \label{eq:wigner}
\end{align}
where $I_{2 \times 2}$ is the identity matrix, $\{\sigma_k\}_{k = 1}^3$ are the Pauli matrices, and $\epsilon_{ijk}$ is the Levi-Civita symbol. Moreover, the Wigner rotation for a quanton that moves over a finite proper time interval can be obtained by iterating the expression for the infinitesimal Wigner rotation \cite{Terashima}, and the spin-$1/2$ representation for a finite proper time can be obtained by iterating the Eq. (\ref{eq:wigner}):
\begin{equation}
    D(W(x, \tau)) = \mathcal{T}e^{\frac{i}{2}\int_0^{\tau} \boldsymbol{\vartheta} \cdot \boldsymbol{\sigma} d\tau'}, \label{eq:time}
\end{equation}
where $\mathcal{T}$ is the time-ordering operator \cite{Terashima}, since, in general, the Wigner rotation varies at different points along the trajectory.

%-------------------------------
\section{Interferometric Visibility in Curved Spacetimes}
\label{sec:inter}
In this section, we will study the behavior of the interferometric visibility of a spin-$1/2$ quanton (or a qubit) in a Mach-Zehnder interferometer in curved spacetimes. Because we are interested in qubits, it is worth pointing out that the motion of spinning particles, either classical or quantum, does not follow geodesics because the spin and curvature couple in a non-trivial manner \cite{Papapetrou}. However, since the deviation from geodetic motion is very small, it can be safely ignored in the cases explored here. 

\subsection{Interferometric Visibility in the Newtonian Limit}
\label{sec:newt}
\begin{figure}[t]
\centering
\includegraphics[scale=0.8]{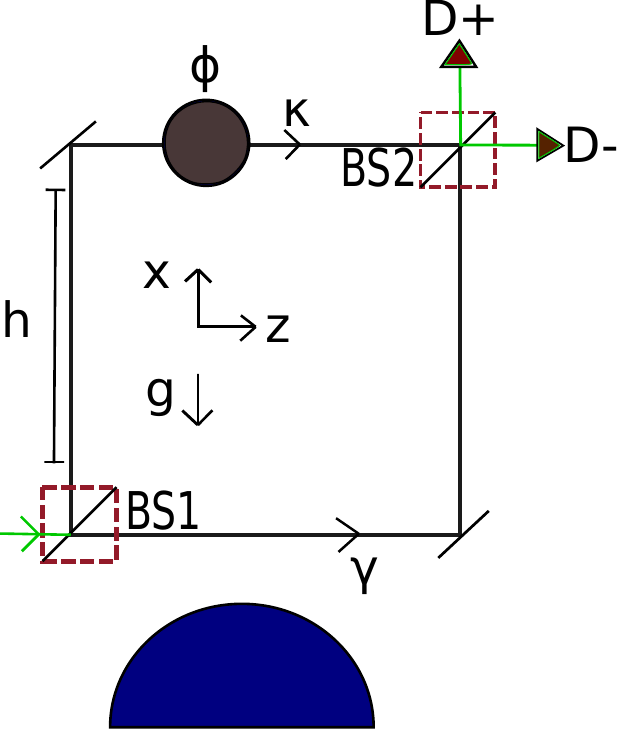}
\caption{Spin-$1/2$ particle in a Mach-Zehnder interferometer in the vicinity of the Earth. The setup consists of two beam splitters $BS1$ and $BS2$, a
phase shifter, which gives a controllable phase $\phi$ to the path $\kappa$, and two detectors $D \pm$. A uniform gravitational field $g$ is oriented anti-parallel to the $x$-direction. }
\label{fig:mach}
\end{figure}

 The Newtonian limit is an approximation applicable to physical scenarios exhibiting: weak gravitation field, objects moving slowly compared to the speed of light, and static gravitational fields \cite{Carroll}. For instance, in the vicinity of the Earth, and in a small region such that the gravitation field is uniform, the spacetime metric can be approximately expressed in the coordinate basis as
 \begin{align}
     ds^2 = - (1 + 2gx)dt^2 + dx^2 + dy^2 + dz^2, \label{eq:newtlim}
 \end{align}
 where $g = GM/R^2$ denotes the value of the Earth's gravitational acceleration in the origin of a laboratory frame (x = 0), which is at distance $R$ from the centre of the Earth and the coordinate $x$ measures the different heights in the gravitational field, as in Fig. \ref{fig:mach}. Considering an interferometric setup where a spin-$1/2$ particle goes through a Mach-Zehnder interferometer, we assume that the quanton is moving with a constant velocity $u^{\mu} = dx^{\mu}/d\tau$. Under this assumption, the particle does not move along a geodesic, hence it must be subject to a non-gravitational interaction to maintain the quanton on its path. Therefore, in this case, the Wigner rotation arises from the external interaction as well from the spacetime geometry effects. Moreover, if the spin degree of freedom can be considered as a `clock', according to general relativity, the proper time should evolve differently along the two arms of the interferometer in the presence of gravity. Because of Bohr's complementarity principle, the interferometric visibility will decrease by an amount given by the which-way information accessible from the final state of the spin, which gets entangled with the path of the quanton, as showed in Ref. \cite{Zych}. The separation between the horizontal arms of the paths $\gamma, \kappa$ of the interferometer is $h$. 

 Here, we will derive the same effect using the unitary representation of the local Lorentz transformation, once Eq. (\ref{eq:unit}) expresses how the spin of the quanton rotates and couples with its momentum as the particle moves along its world line in the interferometer. However, it is worth pointing out that the effect is beyond the linearised regime (first order correction of the metric), as showed by the authors in Ref. \cite{Zych}. In order to calculate the Wigner rotation, we consider the following tetrad field
 \begin{align}
     & e^0_{\ t}  = (1 + 2gx)^{1/2}, \ \ e^1_{\ x}  = e^2_{\ y} = e^3_{\ z} = 1, 
 \end{align}
 and all the other components are zero. Also, only nonzero components will be shown from now on. The inverse of these elements are given by 
\begin{align}
    & e_0^{\ t} = (1 + 2gx)^{-1/2}, \ \ e_1^{\ x}  = e_2^{\ y} = e_3^{\ z} = 1, 
\end{align}
This vierbein represents a static local frame at each point. In addition, at each point, the $0-, 1-, 2-,$ and $3-$axes are parallel to the $t, x, y,$ and $z$ directions, respectively. The components of velocity in the local frame are
\begin{align}
    & u^0 = - u_0 = \sqrt{1 + u^2}, \\
    & u^1 = u_1 = u_x,  \ \ \ u^3 = u_3 = u_z, 
\end{align}
 where $u^2 = u^2_x +  u^2_z$ and $u^y = 0$ since the quanton is restricted to move in the $x-z$ plane. It is worth pointing out that the quanton does not move in $x$ and $z$ directions at the same time. Therefore, the non-zero components of the acceleration due to non-gravitational external interaction is given by:
 \begin{align}
     & a^0 = \frac{g u_x \sqrt{1 + u^2}}{1 + 2gx}, \ \ \ a^1 = \frac{g(1 + u^2)}{1 + 2gx} \label{eq:acce}.
 \end{align}
 Meanwhile, the change of the local frames along the world line is characterized by $ \chi^{a}_{\ b} :=  - u^{\nu}(x) \omega_{\nu \ b}^{\ a}(x)$, which has only one non-zero component given by
 \begin{align}
     \chi^0_{\ 1} = -\frac{g(1 + u^2)^{1/2}}{1 + 2gx}.
 \end{align}
Using the above equations, the non-zero infinitesimal Lorentz transformations $\lambda^a_{\ b}$ can be obtained, and a straightforward calculation similar to those in Ref. \cite{Dai} shows that
\begin{align}
    & \lambda^{0}_{\ 1} = \frac{g (1 + u^2)^{1/2}u^2_z}{1 + 2gx}, \\
    & \lambda^{0}_{\ 3} = - \frac{g (1 + u^2)^{1/2}u_x u_z}{1 + 2gx}, \\
    & \lambda^{1}_{\ 3} = - \frac{g (1 + u^2) u_z}{1 + 2gx},
\end{align}
which expresses that the change in the local frame consists of a boosts along the $1$ and $3$-axis and a rotation about the $2$-axis. It is noteworthy that $\delta p^a = \lambda^a_{\ b}p^b d \tau = 0 $, which is consistent with the assumption that the particle follows straight paths in the interferometer.

Therefore, the Wigner rotation corresponds to a rotation over the 2-axis and is given by:
\begin{align}
      \vartheta^{1}_{\ 3} & = \lambda^{1}_{\ 3} + \frac{\lambda^{1}_{\ 0}p_3 - \lambda_{30}p^1}{p^0 + m} = - \frac{g p_z \sqrt{p^2 + m^2}}{(1 + 2gx) m^2},
\end{align}
where $p = \sqrt{p^2_x + p^2_z}$. One can see that the Wigner rotation depends on $g$, if $g = 0$ there is no rotation. It also depends on the height $x$. Besides, when the particle is moving in the $x$-direction, then $p = p_x$, as well when the particle is moving in the $z$-direction, then $p = p_z$. Now, let us suppose that the initial state of the quanton before $BS1$ in Fig. \ref{fig:mach} is given by
\begin{align}
    \ket{\Psi_{i}} = \ket{p_i} \otimes \ket{\tau_i } = \frac{1}{\sqrt{2}}\ket{p_i} \otimes (\ket{\uparrow} + \ket{\downarrow}),
\end{align}
with the local quantization spin axis along the 1-axis. Right after the BS1, we have a coherent superposition of the paths $\gamma$ and $\kappa$ states such that
\begin{align}
     \ket{\Psi} = \frac{1}{2}(\ket{p_{\gamma}} + i\ket{p_{\kappa}}) \otimes (\ket{\uparrow} + \ket{\downarrow}).
\end{align}
Therefore, the state before the $BS2$ is affected by the gravitational field and is given by
\begin{align}
   U(\Lambda)\ket{\Psi} =& \frac{1}{2}\ket{p_{\gamma}}\otimes D(W(\gamma))(\ket{\uparrow} + \ket{\downarrow}) \nonumber \\ &  + \frac{ie^{i\phi}}{2}\ket{p_{\kappa}} \otimes D(W(\kappa))(\ket{\uparrow} + \ket{\downarrow}),  
\end{align}
where $W(\eta)$ is the total Wigner rotation due to the path $\eta = \gamma, \kappa$. Besides, we choose to maintain the notation $\ket{p_{\eta}}$ for the momentum states before $BS2$, because the velocity of the quanton is constant along the paths. Now, the spin-$1/2$ representation of the Wigner rotation for the paths is given by
\begin{equation}
    D(W(\eta, \tau)) = \mathcal{T}e^{- \frac{i}{2} \sigma_y \int_0^{\tau}  \vartheta^{1}_{\ 3}(\eta) d\tau'}, \ \ \eta = \gamma, \kappa.
\end{equation}
Thus, one can see that $D(W(\eta, \tau))$ depends of the proper time elapsed in each path $\eta = \gamma, \kappa$.

Since parts of the paths are in different heights,  general relativity predicts that the amount of the elapsed proper time is different along the two paths, and therefore the Wigner rotation for each path will be different, which implies that the spin-state will be different and path-information can be accessed. It is worthwhile mentioning that the part of the total Wigner rotation $\Theta(\eta) := \int_0^{\tau}  \vartheta^{1}_{\ 3}(\eta) d\tau'$ of each path $\eta$ that will affect differently the spin-state is due to the horizontal arm of each path of the interferometer. The Wigner rotation due to the vertical route is the same for both paths $\tau, \gamma$ and therefore  it cancels out. Since $\vartheta^{1}_{\ 3}(\eta)$ is constant along both horizontal paths, the time-ordering operator is not needed, and the integration is straightforward. After $BS2$, the states of the momentum are given by $\ket{p_{\gamma}} \to (\ket{p_+} + i \ket{p_-})/\sqrt{2}$ and $\ket{p_{\kappa}} \to (i\ket{p_+} + \ket{p_-})/\sqrt{2}$, and the final state of the quanton is given by
\begin{align}
    \ket{\Psi_f} & = \frac{1}{2\sqrt{2}}(\ket{p_+} + i \ket{p_-}) \otimes e^{- \frac{i}{2}\sigma_y \Theta(\gamma)}(\ket{\uparrow} + \ket{\downarrow}) \nonumber \\
    & + \frac{i e^{i \phi}}{2\sqrt{2}}(i\ket{p_+} + \ket{p_-}) \otimes e^{- \frac{i}{2}\sigma_y \Theta(\kappa)}(\ket{\uparrow} + \ket{\downarrow}). \label{eq:sta1}
\end{align}
Tracing out the spin states in equation (\ref{eq:sta1}) gives the detection probabilities
\begin{align}
    P_{\pm} = \frac{1}{2} \mp \frac{1}{2}\cos(\frac{\Theta(\kappa) - \Theta(\gamma)}{2}) \cos \phi. \label{eq: prob}
\end{align}
When the controllable phase shift $\phi$ is varied, the probabilities $P_{\pm}$ oscillate with amplitude $\mathcal{V}$, which defines the visibility of the interference pattern:
\begin{align}
    \mathcal{V} := \abs{\frac{\max_{\phi} P_{\pm} - \min_{\phi} P_{\pm}}{\max_{\phi} P_{\pm} + \min_{\phi} P_{\pm}}},
\end{align}
where we included the absolute value because the cosine term of the Wigner angles can be positive or negative. Without the entanglement between the internal degree of freedom and the momentum, the expected visibility is always maximal, i.e., $\mathcal{V} = 1$. Whereas, in the case of Eq. (\ref{eq: prob}) it reads \cite{Zych}
\begin{align}
    \mathcal{V} & = \abs{\braket{\tau_{\kappa}}{\tau_{\gamma}}} = \abs{\expval{e^{\frac{i}{2} \sigma_y (\Theta(\kappa) - \Theta(\gamma)) }}{\tau_i}} \nonumber \\
    & =  \abs{\cos(\frac{\Theta(\kappa) - \Theta(\gamma)}{2})},
\end{align}
where $\ket{\tau_{\eta}}$ is the spin-state in the path $\eta$ before $BS2$. Changing the integration variable of $\Theta(\eta) := \int_0^{\tau}  \vartheta^{1}_{\ 3}(\eta) d\tau'$ to the time coordinate of the laboratory frame, it follows that
\begin{align}
  \Theta(\kappa) - \Theta(\gamma) & =  \int_0^{\tau} (\vartheta^{1}_{\ 3}(\kappa) - \vartheta^{1}_{\ 3}(\gamma))  d\tau' \\ & = \int_0^{\Delta T} (\vartheta^{1}_{\ 3}(\kappa) - \vartheta^{1}_{\ 3}(\gamma))  \frac{d\tau'}{dt}dt \\
  & = \alpha \left(\frac{1}{\sqrt{1 + 2gx}} - \frac{1}{\sqrt{1 + 2g(x+h)}}\right)\Delta T \nonumber,
\end{align}
where $\alpha = g p_z \sqrt{p^2 + m^2}/m^2$ and $\Delta T$  is the time  measured in the laboratory frame, for which the particle travels along its world line throughout the interferometer in a superposition of two trajectories at constant heights \cite{Zych}. Using the approximation $(1 + x)^{-1/2} \approx 1 - x/2$, the visibility of the interference pattern is given by
\begin{align}
    \mathcal{V} = \cos (\frac{\alpha \Delta V \Delta T}{2}), \label{eq:visi}
\end{align}
where $\Delta V = gh$  is the difference in the gravitational potential between the paths. The introduction of the internal degree of freedom and its entanglement with the momentum due to the fact that the Wigner rotation depends on the momentum of the particle results in the change of the interferometric visibility. In this case, the difference of the Wigner rotation for each path can be attributed to the difference between the proper time elapsed in each path, as was already shown in Ref. \cite{Zych}, using a different route. It is noteworthy the similarities between Eq. (\ref{eq:visi}) and  the Eq. (13) of Ref. \cite{Zych}. However, it is worth pointing out that the constant $\alpha$ is not related to the energy gap associated with the internal Hamiltonian. Instead, $\alpha$ is related to the gravitational field $g$ and the external degrees of freedom of the particle. This is due to the fact that in the approach taken here, through the representation of the local Lorentz transformation, one does not need to know the (internal) Hamiltonian of the system and how it couples to the gravitation field. Even though, our result has the same functional form of the result obtained in Ref. \cite{Zych}, the approaches to reach the result are conceptually different.

As well, in Ref. \cite{Zych} the acceleration and deceleration undergone by the particle were assumed to be the same for both paths, thus eliminating possible special-relativistic effects. Here, the same thing happens since $u^x$ and $u^z$ are constants. Now, let us focus on the horizontal paths of the Mach-Zehnder interferometer, which are not geodesics. Therefore, it is necessary to consider a non-gravitational interaction to maintain the particle in such horizontal path, since, in the Newtonian limit, defined by Eq.(\ref{eq:newtlim}), the geodesics are parabolas. The authors in Ref. \cite{Zych} did not discuss this issue, neither gave a formula for the acceleration to maintain the particle in such horizontal straight path. 

In contrast, in our manuscript, we give the equation of the acceleration necessary to maintain the particle in such straight path, the Eq. (\ref{eq:acce}). The acceleration is different for the horizontal part of the $\gamma$ and $\kappa$ paths, as one can see this by Eq.(\ref{eq:acce}), since the paths are in different heights. However, this fact does not imply that the velocity $u^z$ will be different in each path. Actually, it implies that, for both paths to have the same velocity $u^z$, it is necessary apply a different acceleration (due to a non-gravitational interaction) in each path. For instance, the acceleration in the path $\kappa$ will be smaller than the acceleration necessary for the path $\gamma$. It is noteworthy that for non-geodetic motion, it is possible, in principle, to use the acceleration due to the non-gravitational interaction to choose the parameters (such as the velocity) of the orbit. In this case, the Wigner rotation arises from the non-gravitational interaction as well from the spacetime geometry effects, even though if $g = 0$ the Wigner rotation vanishes and there are no effect in the interferometric visibility. Nevertheless, it is possible to separate the different effects, once, from Eq. (\ref{eq:visi}), $\alpha$ is the same for both paths and is related to the momentum of the particle, which in turn is related to the non-gravitational interaction needed to maintain the momentum of the particle constant along the path, while $\Delta V \Delta T$ refers to general relativistic effects. On the other hand, if the horizontal path can be approximate as a geodesic with the same velocity in the Newtonian limit, in this case a non-gravitational interaction is notnecessary, and then the Wigner rotation will be only due to spacetime geometry effects, since Eq. (\ref{eq:infloc}) reduces to $\lambda^{a}_{\ b} = \chi^{a}_{\ b}$. 
\begin{figure}[t]
\centering
\includegraphics[scale=0.6]{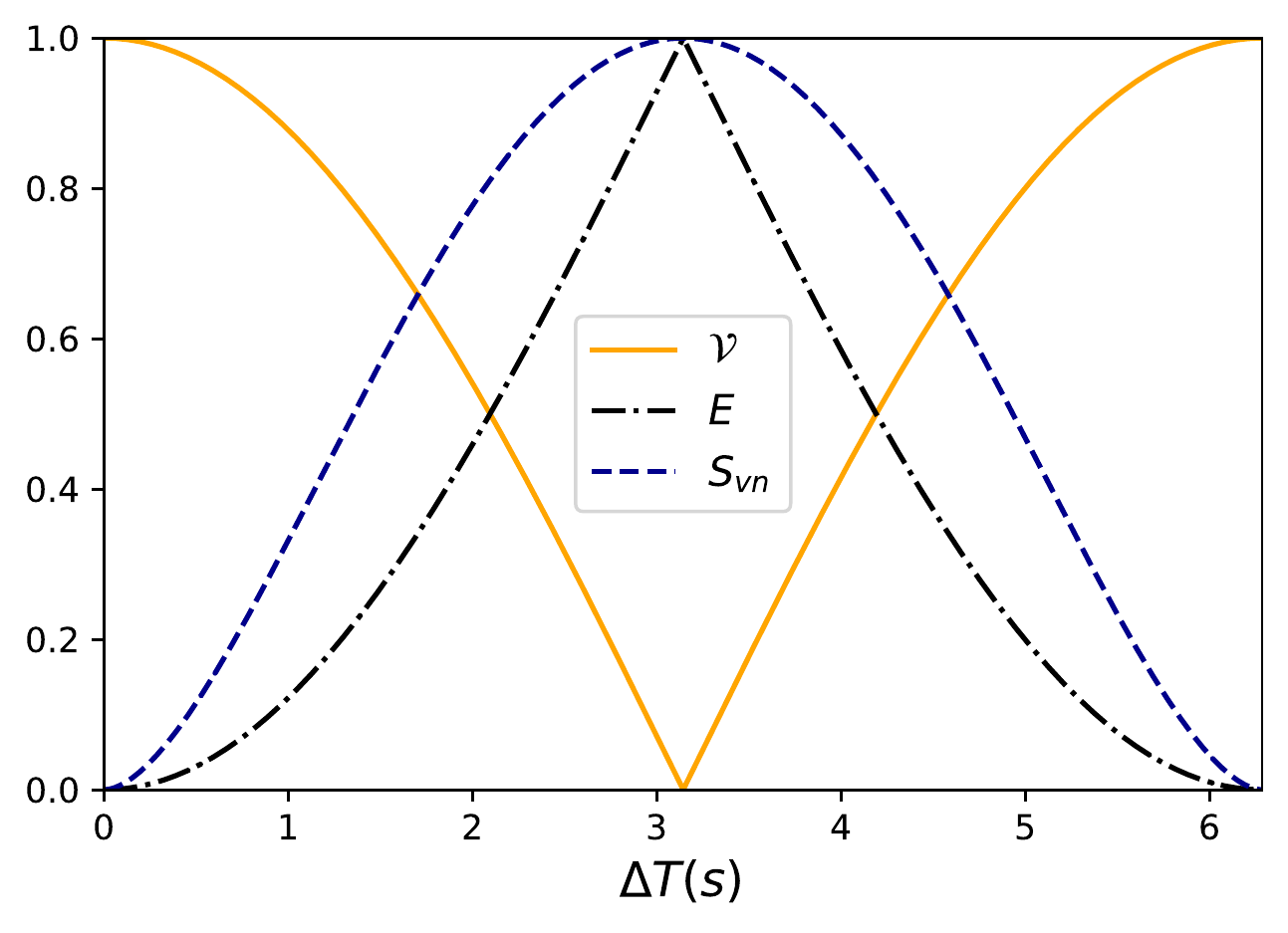}
\caption{Interferometric visibility, $\mathcal{V}$, linear momentum-spin entanglement, $E$, and the spin von Neumann entropy, $S_{vn}(\rho_s)$, as a function of the time $\Delta T$ of the laboratory frame for $\alpha \Delta V = 1s^{-1}$.}
\label{fig:visib}
\end{figure}

Besides, in this case, we can defined $E = 1 - \mathcal{V}$ as a measure of entanglement between the momentum and spin degrees of freedom such that the visibility and entanglement are complementary quantities, as we can see in Fig. \ref{fig:visib}. Also, we plotted the von Neumann entropy of the spin reduced state, $S_{vn}(\rho_s)$, for comparison. In addition, as pointed out in Ref. \cite{Costa}, a clock (evolving internal degree of freedom) with a finite dimensional Hilbert space has a periodic time evolution and thus it is expected that the visibility oscillates  periodically as a function of the difference of the proper times elapsed in the two paths. As we will see, this behavior persists when one studies the interferometric visibility in different spacetime geometries. However, the oscillation is not necessarily due to the difference between the proper times elapsed in each path.

%---------------------
\subsection{Interferometric Visibility in the Schwarzschild Spacetime}
\label{sec:relset}
In this section, we will study the behavior of the visibility of a spin-$1/2$ quanton which is in motion in the Schwarzschild spacetime. In the physical scenario explored in this section, as we will see, the effect of the gravitational field in the interferometric visibility is due to geodetic precession, instead of time dilation. 

The Schwarzschild solution describes the spacetime outside of a static and spherically symmetric body of mass M, which constitutes a vacuum solution. Because of its symmetries, the Schwarzschild metric describes a static and spherically symmetric gravitational field \cite{Hobson}. In the spherical coordinates system $(t, r, \theta, \phi)$, the line element of the Schwarzschild metric is given by
\begin{align}
    ds^2 & = g_{\mu \nu}(x)dx^{\mu} dx^{\nu} \\
    & = -f(r)dt^2 + f^{-1}(r)dr^2 + r^2(d\theta^2 + \sin^2 \theta d\phi^2), \nonumber 
\end{align}
where $f(r) = 1 - r_s/r$, with $r_s = 2GM$ being the Schwarzschild radius. It's straightforward to observe that the metric diverges in two distinct points, at $r = r_s$ and at $r = 0$. However, it is important to distinguish the different nature of both singularities. It is well known that the singularity at $r = r_s$ is not an intrinsic singularity, since it can be shown that all curvature scalars are finite at $r = r_s$, while $r = 0$ is an intrinsic singularity that cannot be removed by changing the coordinate system \cite{Carroll}. To make the Schwarzschild metric reduce to the Minkowski metric, it is possible to choose the following tetrad field
\begin{align}
    & e^0_{\ t}(x) = \sqrt{f(r)}, \ \ \ e^1_{\ r}(x) = \frac{1}{\sqrt{f(r)}} \nonumber, \\
    & e^2_{\ \theta}(x) = r, \ \ \ e^3_{\ \phi}(x) = r \sin \theta,
\end{align}
and all the other components are zero. The inverse of these elements are given by 
\begin{align}
    & e_0^{\ t}(x) = \frac{1}{\sqrt{f(r)}}, \ \ \ e_1^{\ r}(x) = \sqrt{f(r)} \nonumber, \\
    & e_2^{\ \theta}(x) = \frac{1}{r}, \ \ \ e_3^{\ \phi}(x) = \frac{1}{r \sin \theta}.
\end{align}
This vierbein represents a static local frame at each point. Therefore it can used to represent an observer in the associated local frame \cite{Terashima}. In addition, at each point, the $0-, 1-, 2-,$ and $3-$axes are parallel to the $t, r, \theta,$ and $\phi$ directions, respectively.

\begin{figure}[t]
\centering
\includegraphics[scale=0.6]{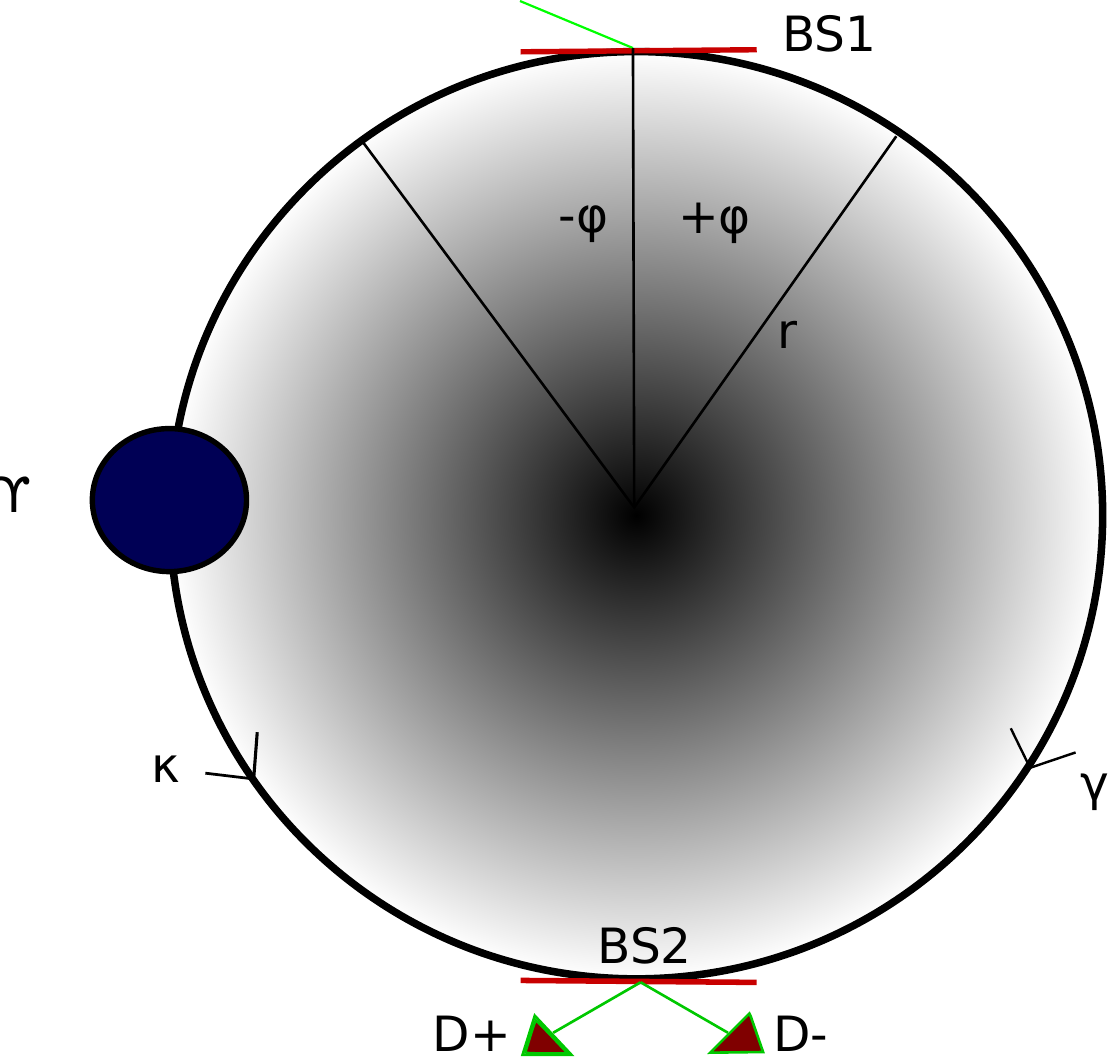}
\caption{Spin-$1/2$ particle in a `astronomical' Mach-Zehnder interferometer. The setup consists of two beam splitters $BS1$ and $BS2$, a phase shifter, which gives a controllable phase $\Upsilon$ to the path $\kappa$, and two detectors $D \pm$. The paths consist of a clockwise and counterclockwise circular geodesic centered in a static and spherically symmetric body of mass M.}
\label{fig:circ}
\end{figure}

Now, let us consider the case of a free-falling test spin-$1/2$ quanton moving around the source of the gravitational field in a superposition of a clockwise and counterclockwise geodetic circular orbit, which play the role of the paths of a Mach-Zehnder interferometer. The four-velocity of these circular geodesics in the equatorial plane, $\theta = \pi/2$, are given by:
\begin{align}
    & u^t = \frac{K}{f(r)}, \ \ \ u^r = 0, \\
    & u^{\theta} = 0, \ \ \ \ \ u^{\phi} = \frac{J}{r^2}, 
\end{align}
where $K, J$ are integration constants related to the energy and angular momentum of the required orbit, respectively, and are given by
\begin{align}
    K = \frac{1 - r_s/r}{\sqrt{1 - \frac{3r_s}{2r}}}, \ \ \ J^2 = \frac{1}{2} \frac{r r_s}{1 - \frac{3r_s}{2r}}.
\end{align}
The energy of the spin-$1/2$ quanton of rest mass $m$ in a circular orbit of radius $r$ is then given by $E = K m$. Furthermore, the value of $J$ implies that the angular velocity is given by 
\begin{equation}
    u^{\phi} = \pm \sqrt{\frac{r_s}{2r^3(1 - \frac{3r_s}{2r})}},
\end{equation}
which means that stable circular geodesic orbits are only possible when $r > \frac{3}{2}r_s$. The non-zero infinitesimal Lorentz transformations in the local frame defined by the tetrad field are given by \cite{Lanzagorta}
\begin{align}
    & \lambda^{0}_{\ 1} = \lambda^{1}_{\ 0} = - \frac{K r_s}{2r^2f(r)}, \\ 
    & \lambda^{1}_{\ 3} = - \lambda^{3}_{\ 1} = \frac{J \sqrt{f(r)}}{r^2},
\end{align}
which corresponds to a boost in the direction of the 1-axes and a rotation over the 2-axis, respectively. While, the four-velocity in the local frame is found to be
\begin{equation}
    u^a = e^a_{\ \mu}(x)u^{\mu} = \left(\frac{K}{\sqrt{f(r)}}, 0, 0, \frac{J}{r}\right).
\end{equation}
Therefore, the Wigner angle that corresponds to the rotation over the 2-axis is given by:
\begin{align}
      \vartheta^{1}_{\ 3}(x)  & = \frac{J \sqrt{f(r)}}{r^2} \left(1 - \frac{K r_s}{2rf(r)}\frac{1}{K + \sqrt{f(r)}}\right).
\end{align}
After the test particle has moved in the circular orbit across some proper time $\tau$, the total angle is given by
\begin{align}
    \Theta & = \int \vartheta^{1}_{\ 3}(x) d \tau = \int \vartheta^{1}_{\ 3}(x) \frac{d \tau}{d \phi} d\phi \\
    & = \frac{\vartheta^{1}_{\ 3}(x)r^2}{J} \Phi, 
\end{align}
since, for a circular orbit, $r$ is fixed and $\vartheta^{1}_{\ 3}(x), K,$ and $J$ are constants. The angle $\Phi$ is the angle traversed by the particle during the proper time $\tau$. It is noteworthy that the angle $\Theta$ reflects all the rotations suffered by the spin of the qubit as it moves in the circular orbit, which means that there are two contributions: The ``trivial rotation'' $\Phi$ and the rotation due to gravity \cite{Terashima}. Therefore, to obtain the Wigner rotation angle that is produced solely by spacetime effects, it' is necessary to compensate the trivial rotation angle $\Phi$, i.e., $\Omega := \Theta - \Phi$ is the total Wigner rotation of the spin exclusively due to the spacetime curvature, which only depends on the radius of the circular geodesic $r$ and on the mass of the source of the gravitational field expressed by $r_s$.

In Fig. \ref{fig:circ}, we represent a spin-$1/2$ quanton in a `astronomical' Mach-Zehnder interferometer. The physical scenario consists of two beam splitters $BS1$ and $BS2$, a phase shifter, which gives a controllable phase $\Upsilon$ to the path $\kappa$, and two detectors $D \pm$. The paths consist of a clockwise and a counterclockwise circular geodesic centered in a static and spherically symmetric body of mass M. The initial state of the quanton, before BS1, is given by $  \ket{\Psi_{i}} = \ket{p_i} \otimes \ket{\tau_i } = \frac{1}{\sqrt{2}}\ket{p_i} \otimes (\ket{\uparrow} + \ket{\downarrow})$, with the local quantization of the spin axis along the 1-axis. Right after BS1, the state is
\begin{align}
    \ket{\Psi} & = \frac{1}{2}\Big(\ket{p_{\gamma};0} + i\ket{p_{\kappa};0})  \otimes  (\ket{\uparrow} + \ket{\downarrow}) \label{eq:state2},
\end{align}
where $\phi = 0$ is the coordinate of the point where the quanton was putted in a coherent superposition in opposite directions with constant four-velocity $u^a_{\pm} = (K/\sqrt{f(r)}, 0, 0, \pm J/r)$. After some proper time $\tau = r^2 \Phi/J$, the particle travelled along its circular paths and the spinor representation of the finite Wigner rotation due only to gravitation effects is given by
\begin{align}
    D(W(\pm \Phi)) = e^{\mp \frac{i}{2}\sigma_2 \Omega}. \ \ \label{eq:wigrot}
\end{align}
Since $\vartheta^{1}_{\ 3}(x)$ is constant along the path, the time-ordering operator is not necessary. Therefore, the state of the quanton in the local frame at the point $\phi = \pi$ before $BS2$ is given 
\begin{align}
U(\Lambda)\ket{\Psi} & = \frac{1}{2} \ket{p_{\gamma};\pi} \otimes e^{-\frac{i}{2}\sigma_y \Omega}(\ket{\uparrow} + \ket{\downarrow}) \nonumber \\
& + \frac{i e^{i \Upsilon}}{2}\ket{p_{\kappa};-\pi}\otimes e^{\frac{i}{2}\sigma_y \Omega}(\ket{\uparrow} + \ket{\downarrow}).  \label{eq:state1}
\end{align}
The detection probabilities corresponding to Eq. (\ref{eq:state1}) after BS2 is given by
\begin{align}
    P_{\pm} = \frac{1}{2}\Big(1 \mp \cos \Omega \cos \Upsilon \Big),
\end{align}
such that the interferometric visibility is given by
\begin{align}
    \mathcal{V} =  \abs{ \cos \Omega}.
\end{align}
In addition, $ \mathcal{V} = C_{l_1}(\rho_{s})$, where $C_{l_1}(\rho_{s})$ is a measure of quantum coherence \cite{Baumgratz} of the spin-state in the interferometer obtained by tracing over the momentum states. It is noteworthy that the proper time elapsed in both paths is the same, since $d \tau = r^2 d \phi/J$. Therefore the influence on the interferometric visibility is due to another general relativistic effect, i.e., the geodetic precession \cite{Misner}. Since the spin is assumed to be parallel transported along the geodesic, it is well known that, when considering an arbitrary closed path, the spin will not return to its initial state due to the curvature effects. The change of the spin state is associated with the spin precession, once that magnitude of any vector is conserved in the geodesic. Here, the spin is in a superposition of a clockwise and a counterclockwise circular path, and therefore the spin precession of each path is given in opposite directions. Thus, the difference between the total Wigner rotation of the paths ends summing up, which affects the interferometric visibility, given that the local Wigner rotation depends on the momentum, which in turn ends up coupling with the spin.

\begin{figure}[t]
    \centering
    \subfigure[$\mathcal{V}, E$ as a function of $r_s/r$.]{{\includegraphics[scale = 0.5]{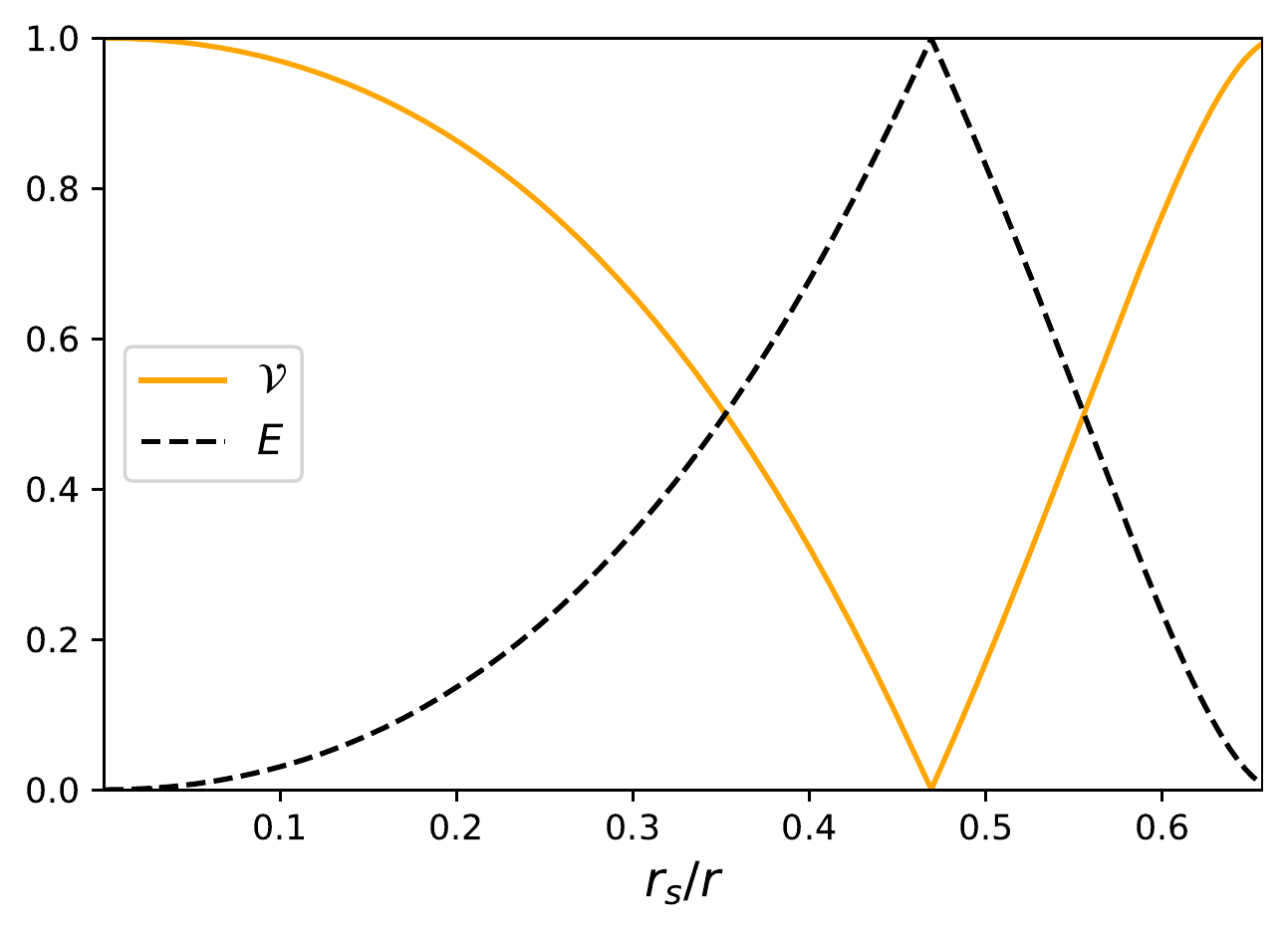}{\label{fig:e}} }}
    \qquad
    \subfigure[The evolution of the visibility along the `astronomical' Mach-Zehnder interferometer for different values of $r$, since $\tau \propto \Phi$.]{{\includegraphics[scale = 0.5]{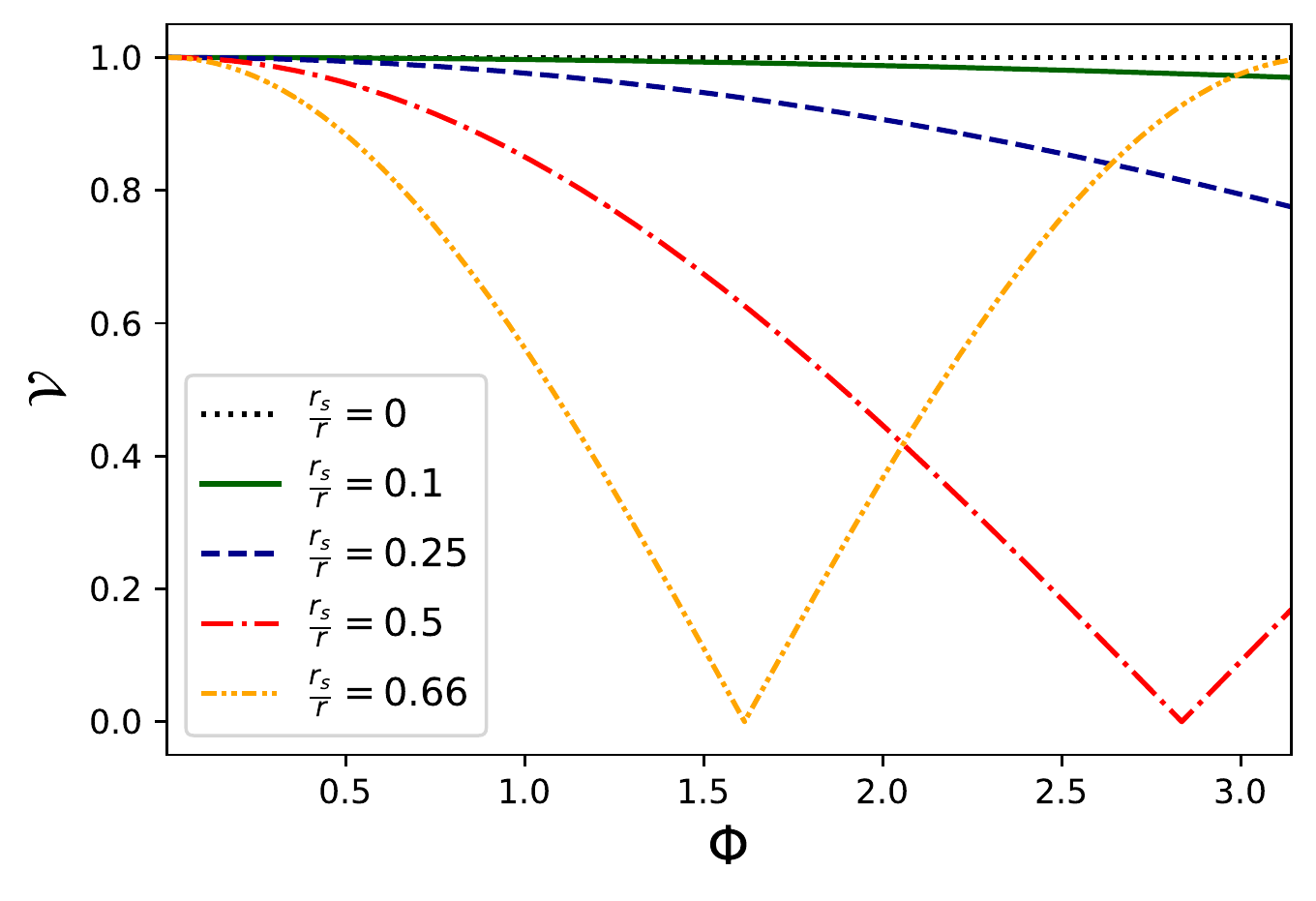}{\label{fig:f}} }}
    \caption{Interferometric visibility and linear momentum-spin entanglement for the Schwarzschild spacetime.}
\end{figure}

Once again, $E = 1 - \mathcal{V}$ can be used to measure the entanglement between the momentum and the spin, as one can see in Fig. \ref{fig:e} and \ref{fig:f}. Specifically, in Fig. \ref{fig:e}, we plotted $\mathcal{V}, E$ as a function of $r_s/r$, when the particle reaches the detectors. Hence, for each value of $r \in (\frac{3}{2}r_s, \infty)$, we have a specific value for  $\mathcal{V}, E$. While, in Fig. \ref{fig:f}, we plotted the `evolution' of the visibility (or the $l_1$-norm quantum coherence) along the interferometer for different values of $r$, since $\tau \propto \Phi$. For instance, when $\mathcal{V}, \Upsilon = 0$, the global state is maximally entangled and is given by
\begin{align}
    \ket{\Psi} =  \frac{1}{\sqrt{2}}\Big( \ket{p_{\gamma}, \uparrow} + i\ket{p_{\kappa}, \downarrow} \Big),
\end{align}
 which implies that the path information is accessible in the internal degree of freedom. Besides, it is possible to extend the orbits closer to $r_s$ by considering non-geodetic circular orbits. In order for the particle to maintain such non-geodetic circular orbit, it is necessary to apply an external radial interaction against gravity and the centrifugal force, allowing the quanton to travel in the circular orbit with the specific angular velocity at a given distance $r$ from the source. In this case, as $r \to r_s$, $\Omega$ varies very rapidly such that, at the event horizon, in the strong field limit $\lim_{r \to r_s} \Omega = - \infty$ \cite{Terashima, Lanzagorta}. This fact will cause the visibility of the quanton to oscillate very rapidly near the Schwarzschild radius. This fact is due to the choice of the tetrad field, and thus the particular static observer, as well as the four-velocity which becomes singular at the horizon. This was noticed also by Terashima and Ueda in Ref. \cite{Terashima}. Therefore, we can conclude that the local static observers, which are not necessarily inertial, attribute a very rapid precession near the horizon.

%----------------------
\subsection{Interferometric Visibility in an Arbitrary Curved Spacetime}
As noticed in Ref. \cite{Costa}, a `clock' (in our case the quanton's spin) with a finite dimensional Hilbert space has a periodic time evolution which causes periodic losses and rebirths of the visibility with increasing time dilation between the two arms of the interferometer. For a clock implemented in a two-level system (which evolves between two mutually orthogonal states), the visibility will be a cosine function. However, as we showed in Sec. \ref{sec:relset}, the effect on the interferometric visibility is due to another general relativistic effect, i.e., the geodetic precession. In this section, we show that by using the unitary representation of the local Lorentz transformation, this effect on the interferometric visibility due to spacetime effects is general and have the same form, i.e., it is a cosine function, provided that we restrict the motion of the quanton to a spacial plane.

\begin{figure}[t]
\centering
\includegraphics[scale=0.6]{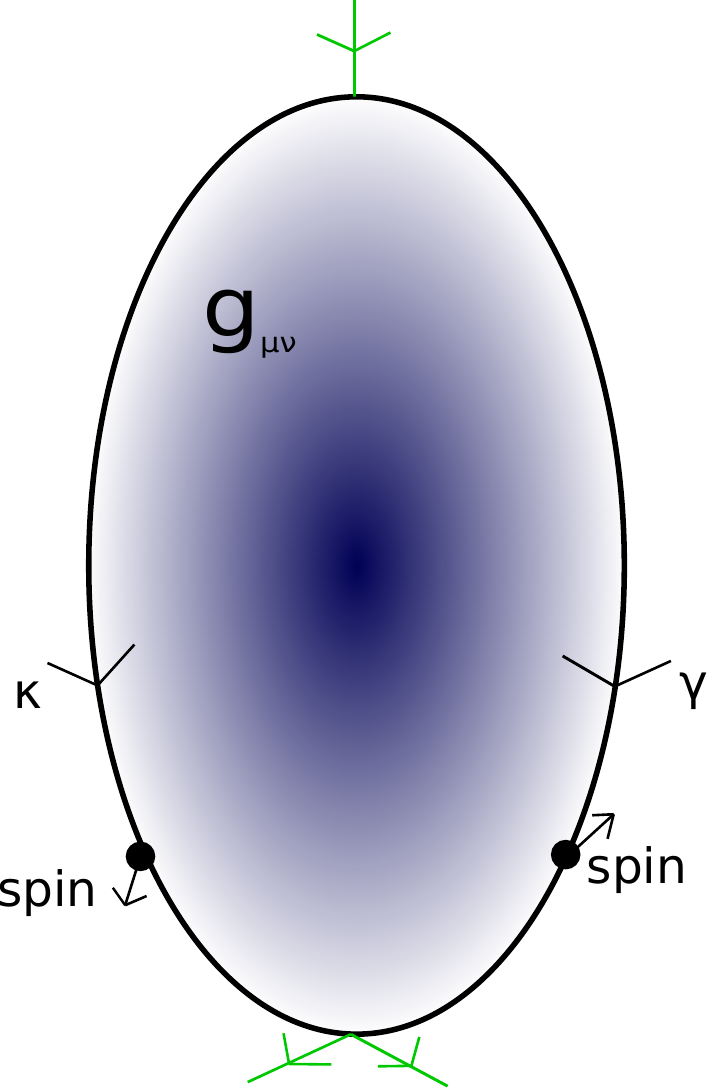}
\caption{An arbitrary curved spacetime $\mathcal{M}$ with metric $g_{\mu \nu}(x)$ where a spin-$1/2$ particle in a superposition of paths $\gamma$ and $\kappa$ is located.}
\label{fig:curv}
\end{figure}

As depicted in Fig. \ref{fig:curv}, let us consider an arbitrary curved spacetime $\mathcal{M}$ with metric $g_{\mu \nu}(x)$, where a spin-$1/2$ particle in a superposition of paths $\gamma$ and $\kappa$ is located. Given that the motion of the quanton is restricted to a two-dimensional spacial plane, it is possible to choose local frames (a tetrad field) such that the Wigner rotation takes place along a single direction. For instance, let us consider the $2$-axis, perpendicular to the plane. In addition, the quantization axis is along the $1$- or $3$-axis located at the plane of motion. Therefore, the corresponding $2$-level unitary representation of the Wigner rotation in the path $\eta = \gamma, \kappa$ is given by
\begin{align}
    D(W(\eta)) = e^{-\frac{i}{2} \sigma_y \int_0^{\tau}  \vartheta^{1}_{\ 3}(\eta) d\tau'} = e^{-\frac{i}{2} \sigma_y \Theta (\eta)}, \ \ \
\end{align}
for paths in which the time ordering operator is not necessary. Therefore, if the initial state of the system before the superposed paths is $\ket{\Psi_{i}} = \frac{1}{\sqrt{2}}\ket{p_i} \otimes (\ket{\uparrow} + \ket{\downarrow}),$ the state of the quanton before the end of the superposed path is given by
\begin{align}
    \ket{\Psi} & = \frac{1}{2}\ket{p_{\gamma}} \otimes e^{-\frac{i}{2} \sigma_y \Theta (\gamma)} (\ket{\uparrow} + \ket{\downarrow}) \nonumber \\
    &  + \frac{1}{2}\ket{p_{\kappa}} \otimes e^{-\frac{i}{2} \sigma_y \Theta (\kappa)} (\ket{\uparrow} + \ket{\downarrow}) \label{eq:state3}
\end{align}
and the interferometric visibility related to state (\ref{eq:state3}), seen by a hovering observer located at point where the overlapping paths meet, is given by
\begin{align}
    \mathcal{V} & = \abs{\expval{e^{\frac{i}{2}\sigma_y (\Theta(\kappa) - \Theta(\gamma)}}{\tau_{ini}}} \nonumber \\
    & = \abs{\cos(\frac{ (\Theta(\kappa) - \Theta(\gamma)}{2})}\nonumber \nonumber  \\
    & = \abs{\cos(\frac{ \int_0^{\tau}(\vartheta^{1}_{\ 3}(\kappa) -\vartheta^{1}_{\ 3}(\gamma))d \tau'}{2})} \label{eq:param},
\end{align}
where $\ket{\tau_{ini}} = (\ket{\uparrow} + \ket{\downarrow})/\sqrt{2}$. One can see again the form of a cosine function for the visibility, as suggested in Ref. \cite{Costa}. However, the effect on the visibility is not necessarily given by the time dilation between the two superposed paths, since it also depends on the Wigner rotation of each path. As we showed before, it can be due to another general relativistic effect, for instance, the geodetic precession. It is noteworthy that the effect on the visibility can be solely due to general relativistic effects if the superposed paths $\gamma$ and $\kappa$ are geodesics. In this case, the Wigner rotation will be only due to spacetime geometry effects, since Eq. (\ref{eq:infloc}) reduces to $\lambda^{a}_{\ b} = \chi^{a}_{\ b}$. On the other hand, if the paths are not geodesics, it is not guaranteed that the effect on the visibility will be only due to spacetime geometry, but as well due to a non-gravitational interaction. Besides, it is always possible to reparametrize the integral in Eq.(\ref{eq:param}) in terms of another convenient coordinate variable. These results can be easily generalized to states with spread momentum (i.e., a wave packet). Therefore, the gravitational field (the spacetime geometry) entangles the momentum and spin of the quanton, such that the momentum state can be, in principle, accessed via the internal degree of freedom, thus affecting the interferometric visibility. Depending on the physical scenario, the effect of the spacetime geometry can manifest in different ways. For instance, in the form of time dilation as in Sec. \ref{sec:newt}; as well in the form of geodetic precession as in Sec. \ref{sec:relset}. On the other hand, if we consider an stationary and axisymmetric spacetime, as the Kerr spacetime, such effect on the interferometric visibility can be due to the well known gravitomagnetic clock effect \cite{Cohen, Tartag}.

At last, if the superposed path is not restricted to a plane, but the Wigner angle does not depend on the paths, we can see that
\begin{align}
    \mathcal{V} & = \abs{\expval{e^{\frac{i}{2}\boldsymbol{\vartheta} \cdot \boldsymbol{\sigma} (\int_{\kappa} -  \int_{\gamma}) d\tau}}{\tau_{ini}}}\nonumber \\
    & = \abs{\expval{e^{\frac{i}{2}\boldsymbol{\vartheta} \cdot \boldsymbol{\sigma}\Delta \tau}}{\tau_{ini}}}\nonumber \\
    & = \abs{\sum_{n = 0}^{\infty}(\frac{i\Delta \tau}{2})^n \expval{(\boldsymbol{\vartheta} \cdot \boldsymbol{\sigma})^n}} \label{eq:moment},
\end{align}
i.e., the influence in the visibility is fully described by the moments $\expval{(\boldsymbol{\vartheta} \cdot \boldsymbol{\sigma})^n} := \expval{(\boldsymbol{\vartheta} \cdot \boldsymbol{\sigma})^n}{\tau_{ini}}$. Expanding Eq. (\ref{eq:moment}) up to second order in $\Delta \tau$, it follows that
\begin{align}
    \mathcal{V} & =  \sqrt{1 - \Big(\frac{\Delta \tau \Delta(\boldsymbol{\vartheta} \cdot \boldsymbol{\sigma})}{2}\Big)^2} \nonumber \\ & \approx 1 - \frac{1}{2}\Big(\frac{\Delta \tau \Delta(\boldsymbol{\vartheta} \cdot \boldsymbol{\sigma})}{2}\Big)^2, \label{eq:decoh}
\end{align}
which is similar to an expression already obtained in Ref. \cite{Costa}, where, in this case,  $\Delta(\boldsymbol{\vartheta} \cdot \boldsymbol{\sigma}) = \expval{(\boldsymbol{\vartheta} \cdot \boldsymbol{\sigma})^2} -  \expval{\boldsymbol{\vartheta} \cdot \boldsymbol{\sigma}}^2$ is variance of $\boldsymbol{\vartheta} \cdot \boldsymbol{\sigma}$, and not the variance of the internal Hamiltonian of the system. From Eq. (\ref{eq:decoh}), we can see that the initial decrease (or decoherence) of the visibility due do spacetime effects is a general phenomena. Finally, for the most general case, i.e., in situations wherein one cannot disregard the time ordering operator, it is possible to construct a Dyson series for $D(W(\eta))$ from Eq. (\ref{eq:time}). The benefit of the approach taken here, through the representation of the local Lorentz transformation, is that we do not need to know the (internal) Hamiltonian of the system and how it couples to the gravitation field, as was the case in Refs. \cite{Zych, Costa}. In contrast, in our approach what is needed is the calculation of the Wigner rotation.%, which is a mechanic and straightforward procedure.  

%---------------------
\section{Conclusions}
\label{sec:con}
In this article, we extended the work of Ref. \cite{Zych} for different spacetime geometries. First, we derived the same effect reported in Ref. \cite{Zych} using the unitary representation of the local Lorentz transformation in the Newtonian Limit. In addition, we showed that the effect in the interferometric visibility due to gravity persists in different spacetime geometries. However, the oscillation is not necessarily due to the difference between the proper times elapsed in each path. For instance, in the setup constructed in Sec. \ref{sec:relset}, the influence on the interferometric visibility is due to another general relativistic effect, i.e., geodetic precession. Besides, we showed that by using the unitary representation of the local Lorentz transformation, this behavior of the interferometric visibility in an arbitrary spacetime is general, provided that we restrict the motion of the quanton to a two-dimensional spacial plane. However, we did not took into account the spin-curvature coupling, which is relevant in the case of supermassive compact objects and/or ultra-relativistic test particles. Finally, we believe that our work helps in the understanding of how the interferometric visibility  of a quantum system is affected due to general relativistic effects as well it opens the possibility for different studies. For instance, it would be interesting to study the behavior of the interferometric visibility in the Kerr geometry, where one takes into account the frame dragging.

\begin{acknowledgments}
This work was supported by the Coordena\c{c}\~ao de Aperfei\c{c}oamento de Pessoal de N\'ivel Superior (CAPES), process 88882.427924/2019-01, and by the Instituto Nacional de Ci\^encia e Tecnologia de Informa\c{c}\~ao Qu\^antica (INCT-IQ), process 465469/2014-0.
\end{acknowledgments}

%--------------------


\begin{thebibliography}{10}
\bibliographystyle{apsrev4-1}

\bibitem{Bose} S. Bose, A. Mazumdar, G. W. Morley, H. Ulbricht, M. Toro$\check{s}$, M. Paternostro, A. A. Geraci, P. F. Barker, M. S. Kim, and G. Milburn, Spin Entanglement Witness for Quantum Gravity, Phys. Rev. Lett. 119, 240401 (2017).
\bibitem{Chiara} C. Marletto and V. Vedral, Gravitationally Induced Entanglement between Two Massive Particles is Sufficient Evidence of Quantum Effects in Gravity, Phys. Rev. Lett. 119, 240402 (2017).
\bibitem{Carlo} M. Christodoulou and C. Rovelli, On the possibility of laboratory evidence for quantum superposition of geometries, Phys. Lett. B 792, 64 (2019).
\bibitem{Howl} R. Howl, R. Penrose, and I. Fuentes, Exploring the unification of quantum theory and general relativity with a Bose-Einstein condensate, New J. Phys. 21, 043047 (2019).
\bibitem{Futamase} S. Wajima, M. Kasai, and T. Futamase, Post-Newtonian effects of gravity on quantum interferometry, Phys. Rev. D 55, 1964 (1997).
\bibitem{Zych} M. Zych, F. Costa, I. Pikovski, and $\check{C}$. Brukner, Quantum interferometric visibility as a witness of general relativistic proper time, Nat Commun. 2, 505 (2011).
\bibitem{Magdalena} M. Zych, F. Costa, I. Pikovski, T. C. Ralph, and $\check{C}$. Brukner, General relativistic effects in quantum interference of photons, Class. Quantum Grav. 29, 224010 (2012).
\bibitem{Brodutch} A. Brodutch, A. Gilchrist, T. Guff, A. R. H. Smith, and D. R. Terno, Post-Newtonian gravitational effects in quantum interferometry, Phys. Rev. D 91, 064041 (2015).
\bibitem{Costa} M. Zych, I. Pikovski, F. Costa, and $\check{C}$. Brukner, General relativistic effects in quantum interference of "clocks", J. Phys.: Conf. Ser. 723, 012044 (2016).

%\bibitem{Schrodinger} E. Schr\"odinger, Discussion of probability relations between separated systems, Math. Proc. Camb. Phil. Soc. 31, 555 (1935). 
%\bibitem{Rosen} A. Einstein, B. Podolsky, N. Rosen, Can Quantum-Mechanical Description of Physical Reality Be Considered Complete? Phys. Rev. 47, 777 (1935).
%\bibitem{Bell} J. S. Bell, On the Einstein Podolsky Rosen paradox, Physics Physique Fizika 1, 195 (1964).
%\bibitem{Popescu} S. Popescu, Bell's inequalities versus teleportation: What is nonlocality?, Phys. Rev. Lett. 72, 797 (1994).
%\bibitem{Preskill} J. Preskill, Quantum information and physics: Some future directions, J. Mod. Opt. 47, 127 (2000).
%\bibitem{Bennett} C.H. Bennett, G. Brassard, C. Cr\'{e}peau, R. Josza, A.
%Peres, W.K. Wooters, Teleporting an unknown quantum state via dual classical and Einstein-Podolsky-Rosen channels, Phys. Rev. Lett. 70, 1895 (1993).
\bibitem{Terno} A. Peres and D. R. Terno, Quantum Information and Relativity Theory, Rev. Mod. Phys. 76, 93 (2004).
%\bibitem{Czachor} M. Czachor, Einstein-Podolsky-Rosen-Bohm experiment with relativistic massive particles, Phys. Rev. A 55, 72 (1997).
\bibitem{Adami} R. M. Gingrich and C. Adami, Quantum entanglement of moving bodies, Phys. Rev. Lett. 89, 270402 (2002).
\bibitem{Ueda} H. Terashima and M. Ueda, Relativistic Einstein-Podolsky-Rosen correlation and Bell's inequality, Int. J. Quantum Inf. 01, 93 (2003).
\bibitem{Milburn} P. M. Alsing and G. J. Milburn, Lorentz invariance of entanglement, Quantum Inf. Comput. 2, 487 (2002). 
\bibitem{Peres} A. Peres, P. F. Scudo, and D. R. Terno, Quantum entropy and special relativity, Phys. Rev. Lett. 88, 230402 (2002).
\bibitem{Palge} V. Palge and  J. Dunningham,  Entanglement of two relativistic particles with discrete momenta, Ann. Phys. 363, 275 (2015).
\bibitem{Dunningham} V. Palge and J. Dunningham,  Generation of maximally entangled states with sub-luminal Lorentz boost, Phys. Rev. A 85, 042322 (2012).
%\bibitem{Bergou} R. Gingrich, A. Bergou, C. Adami,  Entangled light in moving frames, Phys. Rev. A 68, 042102 (2003).
%\bibitem{Li} H. Li, J. Du,  Relativistic invariant quantum entanglement between the
%spins of moving bodies, Phys. Rev. A 68, 022108 (2003).
%\bibitem{Moon} Y. H. Moon, S. W. Hwang, D. Ahn, Relativistic entanglements of Spin 1/2 particles with general momentum, Prog. Theor. Phys. 112, 219 (2004).
%\bibitem{Lee} D. Lee, E. Chang-Young, Quantum entanglement under Lorentz boost, New J. Phys. 6, 67 (2004).
%\bibitem{Jordan} T. F. Jordan, A. Shaji, E. C. G. Sudarshan, Lorentz transformations that entangle spins and entangle momenta, Phys. Rev. A 75, 022101 (2007).
%\bibitem{Vedral} J. Dunningham, V. Vedral, Entanglement and nonlocality of a single relativistic particle, Phys. Rev. A 80, 044302 (2009).
%\bibitem{Friis} N. Friis, R. A. Bertlmann, M. Huber, B. C. Hiesmayr, Relativistic entanglement of two massive particles, Phys. Rev. A 81, 042114 (2010).
%\bibitem{Nasr} B. Nasr Esfahani, M. Aghaee, Relativistic entanglement for spins and momenta of a massive three-particle system, Int. J. Quantum
%Inf. 09, 1255 (2011).
%\bibitem{Blasone} V. A. S. V. Bittencourt, A. E. Bernardini, M. Blasone, Effects of Lorentz boosts on Dirac bispinor entanglement, J. Phys.: Conf. Ser. 1071, 012001 (2018).
%\bibitem{Vlatko} J. Dunningham, V. Palge, V. Vedral, Entanglement and nonlocality of a single relativistic particle. Phys. Rev. A 80, 044302 (2009).
\bibitem{Alsing} P. M. Alsing and G. J. Milburn, Teleportation with a uniformly accelerated partner, Phys. Rev. Lett. 91, 180404 (2003).
\bibitem{Schuller} I. Fuentes-Schuller and R. B. Mann, Alice falls into a black hole: Entanglement in non-inertial frames, Phys. Rev. Lett. 95, 120404 (2005).
\bibitem{Fuentes} P. M. Alsing, I. Fuentes-Schuller, R. B. Mann, and T. E. Tessier, Entanglement of Dirac fields in non-inertial frames, Phys. Rev. A 74, 032326 (2006).
\bibitem{Terashima} H. Terashima and M. Ueda, Einstein-Podolsky-Rosen correlation in gravitational field, Phys. Rev. A 69, 032113 (2004).
\bibitem{Ueda1} H. Terashima and M.  Ueda, Spin decoherence by spacetime curvature, J. Phys. A: Math. Gen. 38, 2029 (2005).
%\bibitem{Kok} P. Kok, U. Yurtsever, Gravitational decoherence, Phys. Rev. D 68, 085006 (2003).

%\bibitem{Leblond} According with J. -M. L\'evy-Leblond, the term "quanton" was given by M. Bunge. The usefulness of this term is that one can refer to a generic quantum system without using words like particle or wave: J.-M. L\'evy-Leblond, On the Nature of Quantons, Science and Education 12, 495 (2003).

%\bibitem{Bohr} N. Bohr, The quantum postulate and the recent development of atomic theory, Nature 121, 580 (1928).

%\bibitem{Engle} B.-G. Englert, Fringe visibility and which-way information: An inequality, Phys. Rev. Lett. 77, 2154 (1996).
%\bibitem{Durr} S. D\"urr, Quantitative wave-particle duality in multibeam interferometers, Phys. Rev. A 64, 042113 (2001).
%\bibitem{Englert} B.-G. Englert, D. Kaszlikowski, L. C. Kwek, W. H. Chee, Wave-particle duality in multi-path interferometers: General concepts and three-path interferometers, Int. J. Quantum Inf. 6, 129 (2008).
%\bibitem{Maziero} M. L. W. Basso, D. S. S. Chrysosthemos, J. Maziero,  Quantitative wave-particle duality relations from the density matrix properties, Quantum Inf. Process. 19, 254 (2020).
%\bibitem{Tabish} T. Qureshi,  Coherence, interference and visibility, Quanta 8, 24 (2019).
%\bibitem{Marcos} M. Basso, J. Maziero, Complete complementarity relations and its Lorentz invariance, arXiv:2007.14480 (2020).


\bibitem{Wald} R. M. Wald, \textit{General Relativity}, (University of Chicago Press,
Chicago, 1984).
\bibitem{Carroll} S. Carroll, \textit{Spacetime and Geometry: An Introduction to General Relativity}, (Addison-Wesley, Reading, 2004).
\bibitem{Nakahara} M. Nakahara, \textit{Geometry,  Topology  and  Physics}, (Institute  of Physics Publishing, Bristol, 1990).
\bibitem{Misner} C. W. Misner, K. S. Thorne, and J. A. Wheeler, \textit{Gravitation} (WH Freeman, San Francisco, 1973).
\bibitem{Lanzagorta} M. Lanzagorta, \textit{Quantum Information in Gravitational Fields} (Morgan \& Claypool Publishers, California, 2014).

\bibitem{Weinberg} S. Weinberg, \textit{The Quantum Theory of Fields I} (Cambridge University Press, Cambridge, 1995).
\bibitem{Eugene} E. P. Wigner, On Unitary Representations of the Inhomogeneous Lorentz Group, Ann. Math. 40, 149 (1939)
\bibitem{Onuki} Y. Ohnuki, \textit{Unitary Representations of the Poincar\'e group and Relativistic Wave Equations} (World Scientific, Singapore, 1988).

\bibitem{Chandra} S. Chadrasekhar, \textit{The Mathematical Theory of Black Holes}, (Oxford University Press, New York, 1983).
\bibitem{Ryder} L. Ryder, \textit{Introduction to General Relativity}, (Cambridge University Press, Cambrige, 2009).

\bibitem{Salgado} M. Lanzagorta, M. Salgado, Detection of gravitational frame dragging using orbiting qubits, Class. Quantum Grav. 33, 105013 (2016).

\bibitem{Kilian}  P. M. Alsing, G. J. Stephenson Jr., and P. Kilian, Spin-induced non-geodesic motion, gyroscopic precession, Wigner rotation and EPR correlations of massive spin 1/2 particles in a gravitational field, arXiv:0902.1396  [quant-ph] (2009).
\bibitem{Papapetrou} A. Papapetrou, Spinning test-particles in general relativity. I, Proc. R. Soc. London A: Math. Phys. Sci. 209 248 (1951).
\bibitem{Dai} Y. Dai, Y. Shi, Kinetic Spin Decoherence in a Gravitational Field, Int. J. Mod. Phys. D 28, 1950104 (2019).

%\bibitem{Basso} M. L. W. Basso, J. Maziero, Complete complementarity relations for multipartite pure states, J. Phys. A: Math. Theor. 53 465301 (2020).

%\bibitem{Sexl} R. U. Sexl, H. K. Urbantke, \textit{Relativity, Groups, Particles: Special Relativity and Relativistic Symmetry in Field and particle Physics} (Springer, New York, 2001).
%\bibitem{Tung} W.-K. Tung, \textit{Group Theory in Physics} (World Scientific, Philadelphia, 1985).
%\bibitem{Ahn} D. Ahn, H. Lee, Y. H. Moon, S. W. Hwang, Relativistic entanglement and Bell's inequality, Phys. Rev. A 67, 012103 (2003).
%\bibitem{Halpern} F. R. Halpern, \textit{Special Relativity and Quantum Mechanics} (Prentice-Hall, New Jersey, 1968).
%\bibitem{Rhodes}J. A. Rhodes, M. D. Semon, Relativistic velocity space, Wigner rotation and Thomas precession, Am. J. Phys. 72, 943-960 (2004).
\bibitem{Hobson} M. P. Hobson, G. Efstathiou, A. N. Lasenby, \textit{General Relativity: An Introduction for Physicists} (Cambridge University Press, Cambridge, 2006).
\bibitem{Baumgratz} T. Baumgratz, M. Cramer,  M. B. Plenio, Quantifying coherence, Phys. Rev. Lett. 113, 140401 (2014).

\bibitem{Cohen} J. M. Cohen and B. Mashhoon, Standard clocks, interferometry, and gravitomagnetism, Phys. Lett. A 181, 353 (1993).
\bibitem{Tartag} A. Tartaglia, Detection of the gravitomagnetic clock effect, Class. Quantum Grav. 17, 783 (2000).

\end{thebibliography}
\end{document}